\documentclass[]{interact}
\usepackage[caption=false]{subfig}

\usepackage[numbers,sort&compress]{natbib}% Citation support using natbib.sty
\bibpunct[, ]{[}{]}{,}{n}{,}{,}% Citation support using natbib.sty
\usepackage{graphicx}% Include figure files
\usepackage{graphics}% Include figure files
\usepackage{dcolumn}% Align table columns on decimal point
\usepackage{multirow}
\usepackage{bm}% bold math
\usepackage{latexsym}
\usepackage{amssymb}
\usepackage{amsmath}
\usepackage{amsfonts}
\usepackage{layout}
\usepackage{verbatim}
\usepackage{epsfig}
\usepackage{graphicx}
\usepackage{amsbsy}
\usepackage{xcolor}
\usepackage[caption=false]{subfig}
\usepackage[toc]{appendix}
\newcommand{\bea}{\begin{eqnarray*}}
	\newcommand{\eea}{\end{eqnarray*}}
\newcommand{\bne}{\begin{equation*}}
\newcommand{\ede}{\end{equation*}}
\newcommand{\ba}{\arraycolsep 0.3ex \begin{array}{rl}}
\newcommand{\ea}{\end{array}}
\newcommand{\bnen}{\begin{equation}}
\newcommand{\eden}{\end{equation}}
\newcommand{\bean}{\begin{eqnarray}}
\newcommand{\eean}{\end{eqnarray}}
\newcommand{\bsen}{\begin{subequations}}
	\newcommand{\esen}{\end{subequations}}

\newcommand{\bna}{\begin{array}}
	\newcommand{\eda}{\end{array}}
\newcommand{\bnm}{\begin{enumerate}}
	\newcommand{\edm}{\end{enumerate}}

\renewcommand{\vec}[1]{\text{\boldmath{$ #1 $}}}

\newcommand {\pd} [2] {{\partial #1\over\partial #2}}

\def\pz{{\partial}}
\def\Bd{{\bm d}}

\def\Bv{{\bm v}}

\def\Br{{\bm r}}

\def\BXi{{\bm\Xi}}
\def\BL{{\bm L}}

\def\Bk{{\bm k}}
\def\Bq{{\bm q}}

\def\BE{{\bm E}}
\def\BR{{\bm R}}

\def\CR{{\mathcal R}}

\def\CGT{\widetilde{\mathcal G}}
\def\BCR{{\bm\CR}}

\def\RF{{\rm F}}

\def\CD{{\cal D}}

\def\eps{\epsilon}
\def\ve{{\varepsilon}}

\def\Sd{{\textsf d}}
\def\Sod{{\textsf{od}}}
\def\SIC{{\textsf{IC}}}
\def\SLC{{\textsf{LC}}}
\def\Bx{{\bm x}}
\def\By{{\bm y}}

\def\Bp{{\bm p}}
\def\xhat{{\hat\Bx}}
\def\yhat{{\hat\By}}

\def\ie{{\it i.e.\/}}

\def\frac#1#2{{{#1 \over #2}}}

\def\hh{\hskip 0.1em}
\def\frac#1#2{{\textstyle{#1 \over #2}}}
\def\half{\frac{1}{2}}
\def\Half{{1\over 2}}
\def\Tra{\mathop{\textsf{Tr}}}
\def\nd{^{\vphantom{\dagger}}}
\def\ns{^{\vphantom{*}}}

\def\Bze{\boldsymbol{0}}
\def\sket#1{{|  \hh #1 \hh  \rangle}}

\def\sexpect#1#2#3{{\langle \hh #1 \hh | \hh  #2  \hh | \hh #3 \hh \rangle}}

\def\Tra{\mathop{\textsf{Tr}}}

\def\Rep{\textsf{Re}}

\def\fmk{{f\ns_{m\Bk}}}
\def\fnk{{f\ns_{n\Bk}}}

\def\vemk{{\ve\ns_{m\Bk}}}
\def\venk{{\ve\ns_{n\Bk}}}

\def\eamn{{\ve\ns_{\alpha\mu\nu}}}
\def\HOmega{{\hat\Omega}}

\def\LASd{\langle L^\alpha\rangle\nd_\Sd}
\def\LAIC{\langle L^\alpha\rangle\nd_\SIC}
\def\LALC{\langle L^\alpha\rangle\nd_\SLC}
\def\bmapright#1{\ \smash{\mathop{\hbox to 25pt{\rightarrowfill}}\limits_{#1}}\ }

\def\RK#1{\CR^{#1}_\Bk}

\def\vK#1{v^{#1}_\Bk}

\def\figcaptionfont{\fontsize{10}{11}\selectfont}
\def\fignumfont{\fontsize{10}{11}\selectfont\bfseries}
\def\tablecaptionfont{\fontsize{10}{11}\selectfont\leftskip\tabledim\rightskip\tabledim}
\def\tablenumfont{\fontsize{10}{11}\selectfont\bfseries}

\begin{document}
	
\title{Quantum geometry and quantum capacitance in the orbital magneto-electric effect}

\author{\name{James H.~ Cullen,\textsuperscript{a}\thanks{Corresponding author. Email: james.cullen@unsw.edu.au} Daniel P. Arovas,\textsuperscript{b} Roberto Raimondi\textsuperscript{c} and Dimitrie Culcer\textsuperscript{a}}
\affil{\textsuperscript{a}School of Physics, The University of New South Wales, Sydney 2052, Australia; \textsuperscript{b}Department of Physics, University of California at San Diego, La Jolla, California 92093, USA; \textsuperscript{c}Dipartimento di Matematica e Fisica, Universit\`a Roma Tre, Via della Vasca Navale 84, 00146 Roma, Italy}}

\maketitle

\begin{abstract}
The orbital magneto-electric effect (OME), the generation of a steady-state orbital angular momentum (OAM) density by an electric field, is a central component of orbitronics. Nevertheless its microscopic interpretation is complicated by the existence of two qualitatively different contributions: the intra-band \textit{Edelstein term} stemming from the combined effect of the equilibrium OAM and Fermi surface shift, and the inter-band contribution, stemming from {\it Zitterbewegung} induced by the electric field. Here we develop a pedagogical density-matrix formulation of the OME within the framework of the modern theory of Bloch-electron OAM, in which the two contributions naturally emerge from distinct parts of the density matrix, and their distinct quantum geometric structures are clearly seen. In the inter-band contribution, in addition to the local- and itinerant-circulation terms, we show that a band-diagonal displacement, or charge dipole, is present, which is linear in the electric field and governed by the normalized quantum metric. For a tilted massive Dirac cone the {\it Zitterbewegung} OME and the Fermi-surface orbital response occur for orthogonal electric-field orientations, providing a route to separate them experimentally. {The itinerant-circulation contribution is identical to the dipolar one, so in a two-band model the dipole accounts for exactly half of the inter-band response, both in the doped regime and in the insulating gap.} Our recent insights identify quantum-metric engineering as a route for enhancing orbital magneto-electric responses and clarify the relation between modern orbital magnetization, polarization, and orbital torque phenomenology.
\end{abstract}


\section{Introduction}

Interest in orbital dynamics in condensed matter physics has surged in recent years as a result of the rapid rise of orbitronics \cite{Orbitronics-PRL-2005-Shoucheng, Orbitronics-in-action, Rhonald-Rev,OC-Rev-EL-2021-Yuriy,wang2024orbitronics,ado2025magnetic}. The central premise is that Bloch electrons carry orbital angular momentum (OAM), and that this orbital degree of freedom can be generated, transported, and converted into magnetic torques in close analogy with spin \cite{OT-FM-PRB-2021-YoshiChika,OT-OEE-NatComm-2018-Haibo,OT-PRR-2020-Hyun-Woo,OT-NatComm-2021-Kyung-Jin, Exp-OT-PRR-2020, LS-conversion-CP-2021-Byong-Guk,OOS-Cvert-2020-PRL-Mathias, PhysRevResearch.4.033037, OHE-OT-large, L-S-OT-2023,OHE-Binghai,tokura2019magnetic,go2023long,Niels_Orbitalsplitter}. The technological motivation driving progress in orbitronics is powerful: if OAM can be generated efficiently and converted into spin inside a ferromagnet or at an interface, it provides an additional route to the electrical control of magnetization, potentially complementing or surpassing conventional spin-orbit-torque mechanisms \cite{Roadmap-SOT-Review,CI-SOT-RMP-2019-Manchon}. Experiments on light metals and weak spin-orbit materials, notably Ti and Cr, have made this possibility especially attractive \cite{Exp-OHE-Ti-Nat-2023-Hyun-Woo,PhysRevLett.131.156702,PhysRevLett.131.156703,10.1063/5.0106988}. At the same time, the microscopic meaning of OAM in a delocalized Bloch band, and of the currents and densities constructed from it, remains more subtle than for spin.

The subtlety originates in the position operator. In a localized atom the orbital magnetic moment (OMM) may be pictured as a current loop formed by an electron orbiting a nucleus. A Bloch electron is spatially extended, so this simple picture cannot be carried over directly to an infinite crystal. The equilibrium OMM of a Bloch wave packet was determined within the framework of the modern theory of orbital magnetization, where it was shown to be related to the geometry of Bloch wave functions and ultimately to the Berry connection and Berry curvature \cite{GaneshSundaram1999, Mingche-1996-PhysRevB.53.7010,Theroy-OM-PRL-2007-Qian, Resta-PhysRevLett.95.137205, Vanderbilt_2018, Resta-PhysRevResearch.2.023139, Resta-PhysRevB.74.024408, thonhauser2011review}. Two complementary formulations have emerged. In the Bloch formulation the electron is envisaged as a wave packet whose internal circulation and field-induced phase-space corrections contribute to the magnetization. In the Wannier formulation the magnetization separates naturally into local circulation within a unit cell and itinerant circulation associated with motion through the crystal and, in finite samples, with boundary currents \cite{Vanderbilt_2018, thonhauser2011review, malashevich2010theory}. These pictures ultimately yield equivalent expressions for the OMM, but they emphasize different aspects of the physics.

Many recent calculations of orbital densities and currents have also employed the atom-centred approximation (ACA), in which the OAM is identified with the angular momentum of atomic-like orbitals entering the band wave functions \cite{IOHE-Metal-PRB-2018-Hyun-Woo,OHE-PRB-2022-Manchon}. This approximation is natural in materials where the orbital character is strongly localized near atomic centres, and it has successfully reproduced equilibrium orbital magnetization in Fe, Co and Ni \cite{OM-Berry-PRB-2016-Mokrousov, lee2026anatomy}. Nevertheless, the ACA is not identical to the modern-theory, which relies on the expectation value of $\bm L=\frac12(\bm r\times\bm v-\bm v\times\bm r)$ over the whole crystal. Recent work has begun to elucidate the connection between the ACA and the modern theory for the OAM in equilibrium \cite{lee2026anatomy}. However, a systematic relation between the ACA and the modern theory out-of-equilibrium remains an open problem. One broad goal of the present work is to keep the modern-theory operator structure explicit and relate it to non-equilibrium OAM densities.

Two electrical routes towards a non-equilibrium OAM have been considered in the literature. The orbital Hall effect (OHE) is the generation of a transverse orbital current by an electric field \cite{Orbitronics-PRL-2005-Shoucheng,ISHE-IOHE-PRB-2008-Inoue,OHE-PRL-2009-Inoue,OHE-metal-PRM-2022-Oppeneer,OHE-BiTMD-PRL-2021-Tatiana,canonico2020two,PhysRevLett.132.106301-Giovanni,veneri2024extrinsic}, while the orbital magneto-electric effect (OME) is the generation of an OAM density by an electric field \cite{OEE-scalar-potential,thonhauser2011review,OM-metal-PRB-2021-Xiaocong,malashevich2010theory,PhysRevB.102.184404,osumi2021kinetic,OEE-NL-2018-Shuichi,Titov_EdgeOM,lee2024orbital,PhysRevB.107.094106,OEE-NatComm-2019-Peter,OAM-SciRep-2017-Yuriy,park2012orbital,johansson2024theory,he2020giant,Edelstein-PhysRevResearch.3.013275,canonico2025spin,hqsf-v45b,LeivaJohansson2025GeTe,CullenGe2DHoles2026, OEE-SciR-2015-Shuichi}. This latter response has appeared under several partially overlapping names: orbital Edelstein effect (OEE), current-induced orbital polarization (CIOP), orbital galvanic or angular-momentum galvanic response, and field-induced orbital magnetization. In this paper we use OME as the generic term for an electrically induced OAM or orbital magnetization density, and reserve OEE for the Fermi-surface contribution brought about by the shift in the electron distribution. The remainder will be referred to as the inter-band contribution, hence, broadly speaking, OME = OEE + inter-band coherence. 

The distinction is motivated by the different physics underlying the two processes. As we will demonstrate below, in the density-matrix picture the OME has two sides. The first is the OEE: the electric field changes the band-diagonal part of the density matrix, $n_{\bm E}$, and this repopulation is traced with the equilibrium OAM carried by each Bloch state. It is the orbital analogue of current-induced spin polarization, depends on relaxation, and is usually proportional to a transport time $\tau$. The second side is the inter-band or intrinsic OME: the electric field produces a band off-diagonal term in the density matrix $S_{\bm E}$, representing coherence between different bands. This contribution is a Fermi-sea response, does not require a Fermi surface, and can survive in an insulating gap. The present work focuses on this second, intrinsic OME, which we refer to as the inter-band contribution, and reveals the hidden geometric displacement inherent in it.

For clarity, Table~\ref{tab:fieldmap} summarizes the main orbitronic responses relevant to this work. The table also highlights a central issue: OAM densities, orbital currents, and orbital torques are not interchangeable. A current may lead to an accumulation only after boundary and relaxation physics are specified; a density may exert a torque only after orbital-to-spin conversion is understood; and both densities and currents depend sensitively on which definition of Bloch-electron OAM is used.

\begin{table}[tbp]
\centering
\small
\begin{tabular}{@{}p{0.18\columnwidth}p{0.20\columnwidth}p{0.38\columnwidth}p{0.18\columnwidth}@{}}
\hline\hline
Response & Density-matrix component & Physical picture & Character \\
\hline
OEE/CIOP & $n_{\bm E}$ & Fermi-surface shift of an equilibrium OAM texture & Dissipative, $\propto\tau$, Fermi surface \\
Inter-band OME & $S_{\bm E}$ & Inter-band coherence and field-induced displacement & Fermi-sea; can survive in gap \\
OHE & $n_{\bm E}$ and $S_{\bm E}$ & Transverse OAM-current response leading to OAM accumulation & Intrinsic and extrinsic \\
Orbital torque & $n_{\bm E}$ and $S_{\bm E}$ & OAM/OAM current generation followed by orbital-to-spin conversion & Interface/material dependent \\
\hline\hline
\end{tabular}
\caption{A schematic map of the orbitronic responses discussed in this work. The OME is an umbrella term for electrically induced OAM density. The OEE is its Fermi-surface, band-diagonal component, while our focus here is primarily on the inter-band coherence component of the OME.}
\label{tab:fieldmap}
\end{table}

The primary technological motivation for both the OHE and the OME is the orbital torque. In an orbital-torque scenario an electric field first generates an orbital current or orbital density in a nonmagnetic layer; the OAM is then injected into, or generated inside, a magnetic layer; spin-orbit coupling converts part of the OAM into spin; and the resulting spin density exerts a torque on the magnetization through exchange coupling \cite{OT-FM-PRB-2021-YoshiChika,LS-conversion-CP-2021-Byong-Guk,PhysRevResearch.4.033037,OHE-OT-large,L-S-OT-2023,cullen2025giant}. Whereas orbital-to-spin conversion remains far from settled at the microscopic level, the OME is especially appealing in two-dimensional systems because it generates a bulk orbital polarization rather than only an edge accumulation.

Several fundamental questions inhibit the development of a predictive theory of orbital torques and orbital magneto-electric phenomena: \textit{What microscopic mechanism produces the inter-band OME, separate from the OEE? How are the Fermi-surface and Fermi-sea responses distinguished experimentally? What is the relation between modern-theory OAM and the atom-centred OAM used in much of the orbitronics literature? Can an OAM density be generated in an insulating gap, and if so what physical motion does it represent?}

In light of the above, the purpose of the present article is twofold. The first is to provide a pedagogical perspective on the OME, emphasizing key theoretical results and experimental observations while organizing the field in density-matrix language. This organization separates two mechanisms that are often discussed together. In the OEE a pre-existing equilibrium OAM texture is converted into a net OAM density by a Fermi-surface shift \cite{OEE-NL-2018-Shuichi}. In the inter-band OME considered here, the electric field generates inter-band coherence; equivalently, it generates OAM even when the equilibrium OAM is zero for each Bloch state. The quantum geometric structures of the two effects are different. The equilibrium OAM is controlled by the antisymmetric, Berry-curvature-like part of the product of two Berry connections, whereas the field-induced dipole below is controlled by the symmetric, quantum-metric-like part of this product. The second purpose is to accompany this perspective with new results that shed light on the origin of the OAM, its generation by electric fields, and fundamental aspects of OAM dynamics in conductors and insulators. Our central result is that a central mechanism behind the inter-band OME is the formation of a non-equilibrium charge dipole. A static electric field induces inter-band coherence, which shifts the centre of the Bloch charge distribution. The resulting dipole is linear in the electric field, survives in insulators, and is controlled by the normalized quantum metric. Explicitly,
\begin{equation}
\begin{split}
d^\alpha &= eE^\beta \sum_{m,\Bk} \CGT^{\alpha\beta,m}_\Bk f_{\Bk m}\,,\\
\CGT^{\alpha\beta,m}_\Bk &=\sum_{m\ne n}{\CR^{\alpha,mn}_\Bk \CR^{\beta,nm}_\Bk + \CR^{\beta,mn}_\Bk \CR^{\alpha,nm}_\Bk\over \ve\ns_{\Bk m} - \ve\ns_{\Bk n}} \quad.
\label{displacement}
\end{split}
\end{equation}
Here $\CGT^{\alpha\beta,m}_\Bk$ is the normalized quantum metric \cite{kaplan2024unification,PhysRevLett.133.106701}, $\CR^{\alpha,mn}_\Bk=i\langle u_{\Bk m}\,|\,{\partial u_{\Bk n}/\partial k^\alpha}\rangle$ is the Berry connection, $f_{\Bk m}$ is the Fermi-Dirac distribution, and the electron charge is $-e$. Equation~(\ref{displacement}) connects the OME to the modern theory of polarization and to the recently discussed quantum capacitance of insulators \cite{komissarov2024quantum}. It also suggests quantum-metric engineering as a strategy for enhancing orbitronic responses.

To quantify this observation we introduce an effective displacement $\bm\Xi$, which is an auxiliary quantity enabling one to determine the expectation value of the position operator in a generic Bloch electron system. Specifically, $\bm\Xi$ keeps track of the infinitesimal wave-vector off-diagonal structure of the density matrix $\rho$ in a gauge-covariant form. Once this information is retained, electric polarization, orbital angular momentum, and orbital magnetization can all be written as traces of ordinary band-space matrices with $\bm\Xi$. {In equilibrium the band-diagonal sector of $\bm\Xi$ is a two-wave-vector distribution containing derivatives of wave-vector delta functions. Fixing its homogeneous trace chooses the origin, but does not set every diagonal matrix element to zero.} Out of equilibrium, its band-diagonal part has a straightforward physical interpretation: it is the electric-field-induced dipole governed by Eq.~(\ref{displacement}). This is a dc manifestation of inter-band coherence, with a natural interpretation in terms of virtual {\it Zitterbewegung} dynamics, and it is directly related to the quantum capacitance of insulators \cite{komissarov2024quantum}.

\begin{figure}[tbp]
\begin{center}
\includegraphics[trim=0cm 0cm 0cm 0cm, clip, width=0.7\columnwidth]{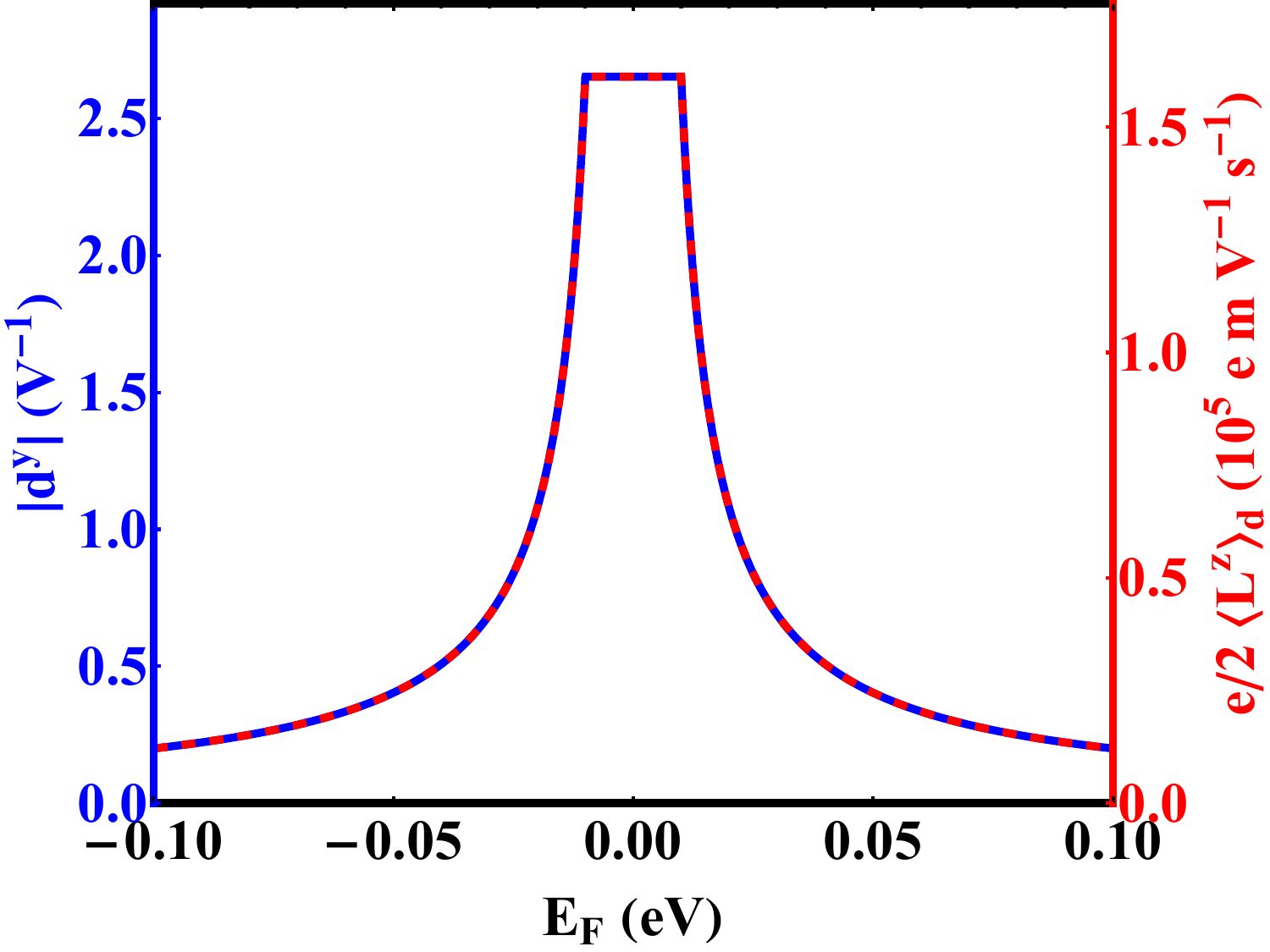}
\caption{\label{Dirac_displacement}
The absolute value of the dipole moment density $|d^y|$ and OMM density $e/2\langle L^z\rangle_{\Sd}$ per unit field for tilted massive Dirac fermions with $\boldsymbol{E}\parallel\yhat$, $\alpha = 4$ eV{\AA}, $m = 0.01$ eV, $v_t = 0.2\, \alpha/\hbar$.}
\end{center}
\end{figure}

Figure~\ref{Dirac_displacement} illustrates this mechanism for a tilted massive Dirac cone. In this model the intrinsic inter-band OME and the Fermi-surface orbital response occur for orthogonal electric-field directions, which provides a simple experimental route for separating them. {In this two-band model the itinerant contribution is identical to the dipolar one at every Fermi energy, so the dipole supplies exactly half of the inter-band OME, in the doped regime and in the insulating gap alike, and explains why an OAM density can be present even without a Fermi surface.} The broader implication is that the OME can be regarded, in suitable systems, as a probe of dipolar dynamics in the quantum geometry of Bloch bands.

The remainder of the article is organized as follows. {Section~2 gives a brief overview of the current state of OME research}. Section~3 then develops the conceptual framework: the density-matrix separation into Fermi-surface and Fermi-sea responses, the effective displacement $\bm\Xi$, the emergence of a quantum-metric dipole, and the decomposition of the OME into dipolar, itinerant-circulation, and local-circulation terms. The tilted Dirac cone is then used as a minimal example that exposes the symmetry distinction between the inter-band OME and the Fermi-surface OEE. The Discussion places these results in the context of modern theory of orbital magnetization and polarization, the spin-Edelstein effect, edge currents, and possible quantization. We close by listing open problems whose resolution is needed for a comprehensive understanding of orbital torques.

\section{OME: Background and status of the field}

The OME is the orbital counterpart of the magneto-electric responses central to spintronics: an applied electric field produces a non-equilibrium angular momentum density. In the spin literature the corresponding response is referred to as a current-induced spin polarization, the inverse spin galvanic effect, or the Edelstein effect. In the orbital literature the terminology is less settled. Closely related calculations and experiments have used the names orbital Edelstein effect, current-induced orbital polarization, kinetic magneto-electric effect, orbital galvanic effect, angular-momentum galvanic effect, and field-induced orbital magnetization. Different names have often emphasized different microscopic pictures: a shift of the Fermi surface in conducting systems, an equilibrium OAM texture in a non-centrosymmetric crystal, an orbital Rashba texture at an interface, or an itinerant orbital magnetization in the modern theory of Bloch electrons. Throughout this paper we use OME for the general electric-field-induced OAM or orbital magnetization density, and OEE for the Fermi-surface contribution obtained by shifting the occupation of Bloch states in momentum space, with the remainder termed the inter-band contribution.

The established theory of the OEE has a straightforward analogy with the spin Edelstein effect. In a non-centrosymmetric conductor an electric field creates a displaced Fermi surface. If the occupied Bloch states carry a momentum-space OAM texture $\bm L_n(\bm k)$, the nonequilibrium occupation $n_{\bm E}$ produces a net orbital moment,
\begin{equation}
    \delta \bm L_{\rm OEE}=\sum_{n\bm k}\bm L_n(\bm k)\, n^{nn}_{\bm E} \quad.
\end{equation}
This is a Fermi-surface effect. It depends on relaxation, is usually proportional to a transport time $\tau$, and vanishes in an insulator at zero temperature. The earliest discussions of current-induced orbital magnetization emphasized chiral crystals and the analogy with solenoids \cite{OEE-NL-2018-Shuichi}. Subsequent work extended the idea to polar metals, noncentrosymmetric antiferromagnets, oxide two-dimensional electron gases, topological and Rashba systems, bilayers, and ferroelectric Rashba semiconductors \cite{OEE-NatComm-2019-Peter,PhysRevB.102.184404,Edelstein-PhysRevResearch.3.013275,osumi2021kinetic,he2020giant,LeivaHenkJohansson2023Bilayer,LeivaJohansson2025GeTe}. A significant part of the OEE literature has employed an atom-centred or projected OAM operator. This is a natural approximation in multiorbital tight-binding models and in first-principles calculations where the orbital character of bands can be resolved in an atomic basis. In that language the OEE is often traced to an orbital Rashba effect: inversion symmetry breaking and orbital hybridization generate a chiral OAM texture, which is then converted into a net orbital polarization by the current. The surface of sp metals, oxide interfaces, Co/Al heterostructures, and crystalline Cu oxide/Cu interfaces provide representative settings in which this language is useful \cite{OAM-SciRep-2017-Yuriy,Exp-OEE-PRL-2022-Jinbo,Nikolaev2024CoAl,Ko2026Cu2O}. Recent first-principles studies of Co/Al, for example, identify a large chiral orbital texture at the interface and argue that the current-induced orbital accumulation can account for large torques in light-element heterostructures \cite{Nikolaev2024CoAl}. Likewise, recent calculations for GeTe using the modern theory of orbital magnetization and semiclassical Boltzmann theory find an orbital Edelstein response roughly an order of magnitude larger than the spin response and largely unaffected when spin-orbit coupling is removed \cite{LeivaJohansson2025GeTe}. These examples show why the OEE has become one of the clearest routes to electrically generated OAM. At the same time, quasi-two-dimensional systems can display orbital responses with a symmetry that is impossible in a strictly two-dimensional Dirac model: in two-dimensional hole gases, for example, transitions between confinement subbands allow an in-plane OME of modern-theory OAM \cite{CullenGe2DHoles2026}.

Modern work on the OAM has shown that the equilibrium orbital magnetic moment of a Bloch electron is not simply the angular momentum of an atomic orbital projected onto a band, but rather a geometric quantity built from the position and velocity operators, and it may be represented in equivalent semiclassical, Bloch, and Wannier formulations \cite{GaneshSundaram1999,Theroy-OM-PRL-2007-Qian,Vanderbilt_2018,Rhonald-Rev}. In the Wannier formulation the orbital magnetization naturally separates into local circulation and itinerant circulation. In the Bloch formulation this physics is encoded in matrix elements of the Berry connection and velocity operator. This distinction matters out of equilibrium because the electric-field response is sensitive to which part of the position operator is retained. The Fermi-surface OEE takes the equilibrium OAM as its starting point and therefore inherits the methodology that was used to determine the OAM texture. By contrast, the inter-band contribution developed below begins from the position operator itself and investigates the electric-field induced dipole structure in the density matrix. In the OME the inter-band sector cannot be reduced to a shifted equilibrium OAM texture.

The Wannier theory of the orbital magneto-electric response identified local and itinerant circulation contributions and clarified the relation to boundary currents in finite samples \cite{malashevich2010theory}. Semiclassical and finite-temperature treatments subsequently expressed electrically induced OAM in terms of band geometry and distribution functions \cite{OM-metal-PRB-2021-Xiaocong,PhysRevB.107.094106}. Kinetic theories of current-induced orbital magnetization in polar systems and topological insulators described the Fermi-surface response in metals \cite{PhysRevB.102.184404,osumi2021kinetic}. More recent work has emphasized edge contributions to itinerant orbital motion \cite{Titov_EdgeOM,lee2024orbital} and ways of generating OEE-like responses from scalar confinement potentials without invoking internal atomic orbital structure \cite{OEE-scalar-potential}. These developments make clear that the OME is not a single mechanism but a family of responses whose microscopic content depends on whether the electric field repopulates bands, mixes bands, displaces charge within the wave packet, or produces an itinerant circulation.

Experimentally, the OME is not always straightforward to isolate. Experiments typically measure torques, magneto-resistance, optical signals, charge currents, or a change in magnetization inferred through a magnetic layer. Several experiments have been interpreted in terms of an orbital Rashba--Edelstein mechanism at oxidized or interfacial Cu structures, orbital torque generation in light-element heterostructures, or conversion between orbital and charge currents \cite{Exp-OEE-PRL-2022-Jinbo,LS-conversion-CP-2021-Byong-Guk,OOS-Cvert-2020-PRL-Mathias,Ko2026Cu2O}. These measurements underline a central difficulty: the observed signal is expected to contain orbital generation, orbital relaxation, orbital-to-spin conversion, and interface transmission. The measured torque depends on the microscopic mechanism by which OAM is converted into spin or lattice angular momentum. This is a primary reason making a clean theoretical separation of the density-matrix channels necessary.

The current status of the field is therefore similar to the early development of spin-orbit torques, but with an additional complication. Spin is a local operator, whereas the modern OAM of Bloch electrons contains the position operator. In the inter-band OME, the electric field produces a static coherence between bands. The associated virtual inter-band motion involves {\it Zitterbewegung}: the field produces an inter-band velocity, and the inverse band splitting supplies the corresponding time scale. The resulting dc displacement is rather different from a non-linear rectification effect. It is linear in the applied field and is the static part of the field-induced inter-band coherence. As shown in the introduction, this displacement is governed by the normalized quantum metric and is closely related to the quantum geometric contribution to capacitance in insulators \cite{kaplan2024unification,PhysRevLett.133.106701,komissarov2024quantum}. This connection provides a different perspective on OME materials -- not merely systems with sizable equilibrium OAM textures, but also possessing a large quantum-metric dipolar response.

The aim of the present work is to make this structure explicit. Since broad reviews of orbitronics have recently appeared elsewhere \cite{Rhonald-Rev, wang2024orbitronics, Orbitronics2DReview, Fukami2025Challenges}, here we focus on the OME alone. Our first goal is pedagogical: to separate the Fermi-surface OEE from the inter-band OME in the density matrix, and to show how the latter is organized by the effective displacement $\bm\Xi$. Our second goal is conceptual: to show that the inter-band OME contains a quantum-metric dipole, interpretable as the dc trace of {\it Zitterbewegung}, and related to quantum capacitance. Our third goal is practical: to identify a minimal tilted Dirac model in which the Fermi-surface and inter-band pieces are selected by orthogonal electric-field directions. The model provides an excellent prototype system which unambiguously brings to light the two sides of the OME by isolating the dipolar OME from the Fermi-surface orbital response.

\section{Theoretical framework}

This section presents the formalism in a pedagogical fashion. We first separate Fermi-surface repopulation from Fermi-sea inter-band coherence, then introduce the effective displacement $\bm\Xi$ as the object that carries the position information normally hidden in the Bloch representation. This makes it possible to identify the quantum-metric dipole before specializing to the OME. Following these general steps we turn our attention to a tilted Dirac cone.

\subsection{Density-matrix response: two contributions to the OME}
The single-particle density matrix $\rho$ encodes both the occupation of Bloch states and the coherence between them. The electric-field response of the density matrix has two qualitatively different pieces: the band-diagonal Fermi-surface response $n_{\bm E}$ and the band off-diagonal inter-band Fermi-sea coherence response $S_{\bm E}$. The first, $n_{\bm E}$, is the distribution shift of the Fermi surface. The second, $S_{\bm E}$, represents coherence between bands. It is the same part of the density matrix that appears in the intrinsic anomalous Hall effect, the spin Hall effect, inter-band polarization in insulators, and the {\it Zitterbewegung} contribution to Dirac transport \cite{Interband-Coherence-PRB-2017-Dimi}.

The expectation value of an operator $\hat O$ is determined by $\Tra[\hat O\rho]$, so the physical content of a linear response can often be read directly from the part of $\rho$ that it probes. This observation is particularly useful for the OME. It naturally separates the response into a Fermi-surface part, which is the conventional OEE, and a Fermi-sea part, which is the intrinsic inter-band OME studied below.

The time evolution of the density matrix is given by the quantum Liouville equation
\begin{equation}
    {\partial \rho\over \partial t}={i\over \hbar}\big[\rho,H\big] \ .
\end{equation}
We use the quantum kinetic formalism based on this equation to determine the non-equilibrium response of the density matrix in an electric field \cite{Interband-Coherence-PRB-2017-Dimi}. The electrostatic potential $H_E=e\bm E\cdot\bm r$ is treated as a perturbation to the band Hamiltonian $H_0$. Here the electron charge is $-e$. To first order in the electric field the kinetic equation is
\begin{equation}
{\pz\rho\ns_\BE\over\pz t}+{i\over\hbar}\big[H_0\,,\,\rho\ns_\BE\big] + J(\rho\ns_\BE)=
-{i\over\hbar}\big[H_\BE,\rho_0\big] \quad ,
\label{eq:qke}
\end{equation}
where $\rho_0$ is the equilibrium density matrix and $J$ captures impurity scattering. In this work $J$ is treated using a simple relaxation time approximation. After disorder averaging, the steady-state response is diagonal in wave vector but not necessarily in the band index. We write
\begin{equation}
\rho_{\bm E}=n_{\bm E}+S_{\bm E}\quad ,
\end{equation}
where $n_{\bm E}$ is band diagonal and $S_{\bm E}$ is band off diagonal. The band-diagonal term $n_{\bm E}$ is the familiar transport response. It describes the electric-field-induced repopulation of states near the Fermi surface and is limited by relaxation. In the simplest relaxation-time approximation,
\begin{equation}
 n^{mm}_{\bm E}= {e\tau\over\hbar}E^\beta\,{\partial f_m\over\partial k^\beta}\quad.
\end{equation}
When this distribution shift is traced with the equilibrium OAM of each Bloch state it gives the OEE. This is the channel calculated in most semiclassical and Boltzmann treatments of current-induced orbital magnetization. It depends on a pre-existing equilibrium OAM texture in momentum space and is generally proportional to $\tau$.

The band off-diagonal term $S_{\bm E}$ is qualitatively different. It represents electric-field-induced coherence between bands and remains finite in the clean limit. In this work we focus on the intrinsic inter-band contribution to $S_{\bm E}$ stemming from the \textit{intrinsic off-diagonal driving term} of Ref.~\cite{Interband-Coherence-PRB-2017-Dimi}, namely
\begin{equation}
S_{\bm E}^{mn}= {e\,\bm E\cdot\BCR^{mn}(f_m-f_n)\over \varepsilon^m-\varepsilon^n}\qquad (m\ne n)\quad,
\label{eq:SE_general}
\end{equation}
which is the same object that gives Berry-curvature corrections to anomalous and spin Hall responses, inter-band polarization in insulators, and {\it Zitterbewegung} contributions in Dirac systems \cite{Interband-Coherence-PRB-2017-Dimi}. The intrinsic Fermi-sea contribution to the OME originates in this inter-band coherence, but because OAM contains the position operator, $S_{\bm E}$ alone is not enough. One must additionally determine the change in the density matrix with respect to an infinitesimal wave-vector difference. This is the role of the effective displacement $\bm\Xi$. This separation is the main conceptual distinction used throughout the paper:
\begin{center}
\begin{tabular}{lll}
\hline\hline
Sector & Response & Physical Mechanism \\
\hline
$n_{\bm E}$ & OEE / CIOP & Equilibrium OAM + Fermi surface shift \\
$S_{\bm E}$ & Intrinsic OME & Inter-band coherence + quantum-geometric displacement \\
\hline\hline
\end{tabular}
\end{center}
Thus the OEE and the inter-band OME represent different parts of the same electric-field response, which can be separated by symmetry, by their transport-time dependence, or by placing the chemical potential in the insulating gap.

For a spatially uniform steady state, $H_0$ and the disorder-averaged density matrix are diagonal in the average wave vector $\bm k$. However, matrix elements of the position operator involve differential operators. Evaluating them requires the dependence of $\rho$ on an infinitesimal wave-vector difference $\bm p$ \cite{Hong-PSHE-PRB, liu2025quantumOHE}. The off-diagonal-in-wave-vector elements of the density matrix are determined by applying the same kinetic equation to $\rho_{\bm k+\bm p/2,\bm k-\bm p/2}$ and then taking the limit $\bm p\rightarrow0$. The dependence of the density matrix on these infinitesimally separated wave vectors is contained in the effective displacement $\bm\Xi$ in the notation used below. The trace of $\bm\Xi$ gives a field-induced dipole, while its contraction with velocity gives the OAM.

\subsection{Effective displacement and dipole operators} 

Determining $n_{\bm E}$ and $S_{\bm E}$ is sufficient for evaluating the expectation values of local operators such as spin, but the OAM requires special care since it contains the position operator. In the Bloch representation the band-diagonal matrix elements of $\bm r$ are differential operators \cite{blount1962formalisms, Position-operator-Rev,resta1998positionop}; they act not only on the band matrix elements of the operator being measured, but also on the wave-vector dependence of the density matrix itself. The starting point must therefore include matrix elements of the density matrix between infinitesimally separated wave vectors.

This physics is captured by introducing the effective displacement $\bm\Xi$. Consider a dipole operator of the form $(1/2)\{A,\bm r\}$, where $A$ is an operator diagonal in wave vector. Its expectation value can be written in the compact form
\begin{equation}
\langle \Bd\ns_A \rangle = \sum_\Bk\sum_{m,n} A^{mn}_\Bk\,\BXi^{nm}_\Bk\equiv \Tra\,[A\,\BXi]\quad,
\label{dipole_op}
\end{equation}
where
\begin{equation}
\BXi^{mn}_\Bk = i\,\bigg( { \pz\rho^{mn}_{\Bk_+ \Bk_-} \over 
\pz\Bp } \bigg)_{\!\Bp\to 0} +
\half\big\{\BCR\ns_\Bk\,,\,\rho\ns_\Bk\big\}^{mn}\quad,
\label{eq:cdod}
\end{equation}
with $\Bk_\pm=\Bk\pm\half\Bp$. In Eq.~(\ref{eq:cdod}) the derivative with respect to $\bm p$ gives the Cartesian components of $\bm\Xi$. The Berry-connection term is essential. It cancels the gauge-dependent part generated by differentiating the Bloch basis, so that $\bm\Xi$ is gauge covariant and the traces in Eq.~(\ref{dipole_op}) are gauge invariant. Details are given in Appendix~A.

The physical content of Eq.~(\ref{eq:cdod}) is simple. A homogeneous density matrix $\rho_{\bm k}$ determines the band occupation as well as coherent band mixing at a fixed wave vector, but it does not by itself contain all position information. The missing information comes from the phase and amplitude structure connecting $\bm k+\bm p/2$ to $\bm k-\bm p/2$. The kinetic equation supplies a constraint equation for this structure. For band off-diagonal elements the constraint fixes $\bm\Xi$ through the relative precession of different bands, whose frequency is $(\varepsilon_m-\varepsilon_n)/\hbar$. {In equilibrium the band-diagonal elements are distributions rather than regular functions of $\bm k$: they contain derivatives of $\delta(\bm k-\bm k')$. Choosing the origin fixes only the homogeneous trace of this kernel and does not set every band-diagonal matrix element to zero.} By contrast, the band-diagonal part of $\bm\Xi_{\bm E}$ is a genuine electromagnetic response. It is the displacement of the charge distribution induced by the electric field relative to the chosen equilibrium origin.

The effective displacement $\bm\Xi$ is the natural language for discussing the inter-band OME. Setting $A=1$ in Eq.~(\ref{dipole_op}) gives the electric dipole; setting $A$ to the velocity gives OAM; and inserting the appropriate current operator gives the orbital-current response. The same density-matrix moment is therefore responsible for polarization, OAM, and orbital transport. In the present paper we make use of it in order to show that the inter-band OME is associated with a quantum-metric dipole, in addition to the internal and itinerant circulation terms familiar from the modern theory of orbital magnetization.

\subsection{Equilibrium OAM as internal circulation}

Calculating $\langle\boldsymbol{r}\rangle$ is equivalent to setting $A=1$ in Eq.~(\ref{dipole_op}). To find ${\bm \Xi}$ in equilibrium the covariant derivative (\ref{eq:cdod}) of the equilibrium density matrix must be calculated. {Its band-diagonal sector is the distributional kernel described above, while its regular band off-diagonal elements are $\Xi^{\alpha,mn}_{0}=\half\,\CR^{\alpha,mn} (f_m + f_n)$ with $m\neq n$. These regular elements generate the familiar internal-circulation contribution to the equilibrium OAM; the complete treatment of the distributional diagonal sector is not needed in the endpoint response below.}

Introducing the effective displacement $\bm\Xi$ enables us to treat position, dipole moments and OAM on the same footing. The OAM operator is defined as $\BL=\half(\Br\times\Bv - \Bv\times\Br)$, with $\Bv$ the velocity operator \footnote{Here we define the operator $\bm L$ without the mass, this is done in order to avoid confusion with the bare and effective electron masses.}. The expectation value of the orbital angular momentum $\langle L^\alpha\rangle$ can be calculated by setting $A$ in (\ref{dipole_op}) to be the velocity and calculating $\ve\ns_{\alpha\mu\nu}\Tra\,[v^\nu\,\Xi^\mu]$. The equilibrium expectation value of $\langle\BL\rangle$ calculated from the band off-diagonal elements of $\BXi$ agrees with the well-known result for the orbital moment in the modern theory \cite{GaneshSundaram1999, Mingche-1995-PhysRevLett.75.1348, Mingche-1996-PhysRevB.53.7010}.

The inter-band matrix elements encode the relative motion of amplitudes in different bands and are the source of the internal circulation captured by the modern theory. Hence, the appearance of these terms in the expectation value of the OAM is unsurprising. This, however, is in many ways the easier part of the problem, since the OAM expectation value in equilibrium involves only inter-band matrix elements of the position operator, and one does not need to consider differential operators in the intra-band elements of $\boldsymbol{r}$.

\subsection{Field-induced displacement and inter-band OME} 

For the orbital magneto-electric effect we are ultimately interested in deviations in the electrons' equilibrium position and OAM brought about by an applied electric field. Both of these require the non-equilibrium correction to $\BXi$ to linear order in the electric field, which we denote by $\BXi_\BE$, which can be found using Eq.~(\ref{eq:SE_general}). In this work we focus on the intrinsic response in $S_{\bm E}$, without including disorder explicitly. {As shown in Appendix B, we retain a finite frequency $z=\hbar(\omega+i0^+)$ and use the symmetric endpoint continuation $\Bk_\pm=\Bk\pm\Bp/2$,
\begin{equation}
H_{0}(\bm k_+)\rho_{\BE,\Bp}(z)-\rho_{\BE,\Bp}(z)H_{0}(\bm k_-)-z\rho_{\BE,\Bp}(z)
=-ieE^\beta(D_\beta\rho_0)_\Bk .
\label{eq:endpoint_qke}
\end{equation}
Here $D_\beta O\equiv\partial O/\partial k^\beta-i[\CR^\beta,O]$ is the band-covariant derivative. The electric source is evaluated at the central wave vector before it is projected between the Bloch states at $\Bk_+$ and $\Bk_-$. Differentiating with respect to $\Bp$ and only then taking $\Bp\to0$ gives
\begin{equation}
[H_0,\Xi^\alpha_\BE(z)]-z\Xi^\alpha_\BE(z)
=-{i\hbar\over2}\{v^\alpha,\rho_\BE(z)\}.
\label{eq:qkeXi}
\end{equation}
After the endpoint accumulation poles have been separated, the part of the regular band off-diagonal displacement that contributes to the OAM is, for $m\ne n$,
\begin{equation}
\left.\Xi^{\alpha,mn}_{\BE,\Bk}\right|_{\rm OAM}
=-{i\hbar\over2(\ve^m-\ve^n)}\{v^\alpha,S_\BE\}^{mn}.
\label{eq:XiE}
\end{equation}
The remaining finite Fermi-surface term is symmetric in the two OAM indices and drops out after contraction with $\epsilon_{\alpha\mu\nu}$. In particular, Eq.~(\ref{eq:qkeXi}) contains no derivative of $\Xi_0$: the distributional equilibrium kernel is correct, but the central-source endpoint continuation does not require a pointwise value for it. We separate the anticommutator in Eq.~(\ref{eq:XiE}) into its repeated-band and all-distinct three-band sectors, which give the itinerant- and local-circulation terms, respectively.}

%\renewcommand{\arraystretch}{1.2}
%\begin{table*}[t!]
\begin{table}[t!]
\centering
%\begin{tabular}{ m{0.5\columnwidth} m{0.5\columnwidth} m{\columnwidth} }
\begin{tabular}{@{}m{0.3\columnwidth}m{0.3\columnwidth}m{0.33\columnwidth}@{}}
\hline\hline
OAM component & Expression & Physical origin\\
\hline
\vspace{1mm}
$\langle L^\alpha\rangle\nd_\Sd$ & $\eps\ns_{\alpha\beta\gamma}\Tra\,
[v^\gamma_\Sd\, \Xi^\beta_{\BE,\Sd}]$ & Electrically induced dipole \\ 
\vspace{1mm}
$\langle L^\alpha\rangle\nd_\SIC$ & {$\big[\eps\ns_{\alpha\beta\gamma}\Tra\,
(v^\gamma_\Sod\, \Xi^\beta_{\BE,\Sod})\big]_{\rm rep}$} & Itinerant circulation \\ 
\vspace{1mm}
$\langle L^\alpha\rangle\nd_\SLC$ & {$\big[\eps\ns_{\alpha\beta\gamma}\Tra\,
(v^\gamma_\Sod\, \Xi^\beta_{\BE,\Sod})\big]_{3{\rm b}}$} & Local circulation \\ 
\hline\hline
\end{tabular}
\caption{\label{tab:OAMcomp}Components of the nonequilibrium OAM and their physical origins. 
{Here ``rep'' and ``3b'' denote, respectively, repeated and all-distinct intermediate band indices.} 
The electrically induced dipole is evaluated as ${\bm d} = \Tra\,\BXi\ns_\BE$, so $\langle L^\alpha\rangle_{\BE,\Sd}$ is straightforwardly related to this dipole.}
\end{table}
%\end{table*}
%\renewcommand{\arraystretch}{1}

The intrinsic inter-band part of the nonequilibrium density matrix is off-diagonal in the band index, but $\BXi$ has both diagonal elements $\BXi_\Sd$ and off-diagonal elements $\BXi_\Sod$. {After the finite-frequency poles are separated, the endpoint equation yields the regular diagonal coefficient as well as the off-diagonal one. The diagonal coefficient agrees with the direct definition in Eq.~(\ref{eq:cdod}): $\BXi_{\BE,\Sd}=\half\{\BCR,S_\BE\}_\Sd$.} The intraband elements of the effective displacement $\BXi_{\BE,\Sd}$ lead to a finite trace and gives us the expected dipole moment in Eq.~(\ref{displacement}). 

The electrically induced dipole in Eq.~(\ref{displacement}) can also be straightforwardly derived using perturbation theory. For this we introduce the electric potential as a perturbation $H\ns_\BE=e\BE\cdot\Br$. Writing the first order in $\BE$ correction to the Bloch wavefunctions as
\begin{equation}
\psi^{(1)}_{n\Bk}(\Br) = \sum_{m\neq n} 
C\ns_{m\Bk}\,\psi^{(0)}_{m\Bk}(\Br) \quad,
\end{equation}
standard first order perturbation theory yields
\begin{equation}
C\ns_{m\Bk}=-e\BE\cdot{\BCR_\Bk^{mn}\over\ve\ns_{m\Bk}-\ve\ns_{n\Bk}}\quad,
\end{equation}
for $m\ne n$. To first order in perturbation theory, the expectation value of the position operator in each state $n\Bk$ is
\begin{equation}\label{Eq:Pert_dipole}
\langle r^\beta\rangle_{n\Bk,\bm E}
=-e E^\alpha\,{\sum_m}'\> {\RK{\alpha,nm}\,\RK{\beta,mn} + \RK{\alpha,mn}\,\RK{\beta,nm}\over\vemk - \venk}\quad,
\end{equation}
from which we can deduce there is a dipole moment induced in state $\sket{\psi\ns_{n\Bk}}$ along the electric field direction. The primed sum in Eq.~(\ref{Eq:Pert_dipole}) indicates that we sum over distinct band indices. Furthermore, by tracing Eq.~(\ref{Eq:Pert_dipole}) with the equilibrium density matrix we recover the dipole density in Eq.~(\ref{displacement}).%3

Equation~(\ref{Eq:Pert_dipole}) constitutes the most direct pedagogical expression of the dipole physics introduced in this work. The numerator is the real, metric part of the inter-band quantum geometry, while the denominator is the inter-band energy separation. Thus a large intrinsic dipole is expected in systems with strong inter-band mixing and small direct gaps, provided symmetry allows the response. The same structure underlies the recently discussed quantum capacitance of insulators: a static electric field changes the charge distribution even when no Fermi-surface carriers are present. In the present problem this displacement is in general not to be interpreted as an electric polarization - rather, when combined with velocity, it becomes part of the OAM response.

The non-equilibrium expectation value of the OAM in response to an electric field can be separated into three terms $\langle L^\alpha\rangle_{\Sd/\SIC/\SLC}$ (see Table \ref{tab:OAMcomp}). The explicit expressions for each contribution to the inter-band OME are
\begin{equation}\label{Eq:Ld}
\LASd =-\eamn\, eE^\beta\sum_\Bk \sum_{m\ne n} \Rep\bigg[\bigg( {\vK{\mu,mm} +
\vK{\mu,nn}\over\vemk-\venk}\bigg)\RK{\nu,mn}\,\RK{\beta,nm}\bigg]\fmk \quad ,
\end{equation}
\begin{align}\label{Eq:LLC}
\LALC&= -\Half\,e E^\beta\,\eamn\sum_\Bk{\sum_{m,n,\ell}}^{\!\prime}\Big(
\RK{\beta,mn}\,\RK{\mu,n\ell}\,\vK{\nu,\ell m} +
\RK{\mu,m\ell}\,\vK{\nu,\ell n}\,\RK{\beta,nm}\\
& \hskip 0.8in  +\vK{\nu,m\ell}\,\RK{\mu,\ell n}\,\RK{\beta,nm} +
\RK{\beta,mn}\,\vK{\nu,n\ell}\,\RK{\mu,\ell m}\Big)
\bigg({\fmk\over\venk - \vemk}\bigg)\quad ,\nonumber
\end{align}
{
\begin{align}\label{Eq:LIC}
\langle L^\alpha\rangle\nd_\SIC =&\ \eps\ns_{\alpha\mu\nu}\,eE^\beta
\sum_{n\ne m}\sum_\Bk\Rep\bigg[\bigg( {\vK{\nu,nn} + \vK{\nu,mm} \over \venk-\vemk} \bigg)
\RK{\mu,nm}\,\RK{\beta,mn}\bigg]\fnk
=\LASd\quad .
\end{align}}
where it is understood that all the Berry connections in these expressions are only off-diagonal in band index. {Whereas the band-diagonal component $\BXi_{\BE,\Sd}$ gives the dipole, the band off-diagonal contraction in Eq.~(\ref{eq:XiE}) separates into a repeated-band contribution $\langle L^\alpha\rangle_\SIC$ and an all-distinct three-band contribution $\langle L^\alpha\rangle_\SLC$. Equation~(\ref{Eq:LIC}) shows that the itinerant term is equal to the dipolar term, although it arises from the band off-diagonal displacement rather than from the charge dipole itself. The local-circulation term in Eq.~(\ref{Eq:LLC}) requires three distinct bands and therefore vanishes in a two-band model. We retain the conventional itinerant/local terminology for these algebraic groupings; their complete sum is independent of this separation.} The contribution $\langle L^\alpha\rangle_\Sd$ in Eq.~(\ref{Eq:Ld}) contains $\BXi\ns_{\BE,\Sd}$ which is solely responsible for the steady-state dipole $\Bd$ in (\ref{displacement}). Hence, $\langle L^\alpha\rangle\ns_\Sd$ represents a steady-state dipole convected along with the electron, generating OAM. This can be seen in Fig.~\ref{Dirac_displacement}, as $\Bd$ and $\langle L^\alpha\rangle\ns_\Sd$ have identical behavior as a function of the Fermi energy. In contrast to the equilibrium OAM, whose standard expression is controlled by the imaginary part of inter-band Berry-connection products, the dipolar contribution to the nonequilibrium OAM is controlled directly by the real, metric part. The remaining terms are also built from inter-band Berry connections and velocities, but only the band-diagonal displacement has the direct interpretation of a quantum-metric charge dipole. Detailed comparisons between our OME results and earlier work \cite{OM-metal-PRB-2021-Xiaocong, malashevich2010theory, PhysRevB.107.094106} are discussed in the Discussion. In principle, our formulae for the OME can be directly applied to any model. However, systems that cannot be captured by simple analytical models, such as Moir\'e superlattices, likely require a detailed analysis \cite{he2020giant}.

\section{Prototype system: tilted Dirac cone}

We consider a massive Dirac cone with a tilt in the $\xhat$ direction, $\hbar v_t k^x$, as a minimal prototype model. We first isolate the inter-band part of the density matrix, which is the contribution calculated in the general theory above, and then compare it with the Fermi-surface OEE in the same model. Two reasons underlie our choice of model. The first is that a tilted massive Dirac cone provides analytical insight into the interplay of inversion symmetry, particle-hole asymmetry, the insulating-gap response, and inter/intra-band mechanisms. The same Hamiltonian structure appears locally in many topological and valley systems, including transition-metal dichalcogenides, graphene-derived semimetals, magnetic topological surfaces, and topological antiferromagnets. The second is that, remarkably, it allows one to distinguish between inter-band and Fermi-surface contributions to the OME by orienting the electric field along different directions, as we shall see below. The model Hamiltonian is expressed as
\begin{equation}
    H =\hbar v_t\,k^x\mathbb{I}+ m\sigma^z+\alpha (k^y\sigma^x - k^x\sigma^y)\quad.
\end{equation}
For this model the electrically induced dipole for the Fermi energy in the conduction band is
\begin{equation}
    \langle y \rangle =-{e E^y(m^2 + 3E_\RF^2)\over 48\pi E_\RF^3}\quad,
\end{equation}
where we take $\boldsymbol{E}\parallel\yhat$. The induced displacement is collinear with the electric field. In the insulating case, with the Fermi energy in the gap, one may take $E_\RF\rightarrow m$, giving $\langle y \rangle =-e E^y/12\pi m$.

\begin{figure}[tbp]
\begin{center}
\includegraphics[trim=0cm 0cm 0cm 0cm, clip, width=0.7\columnwidth]{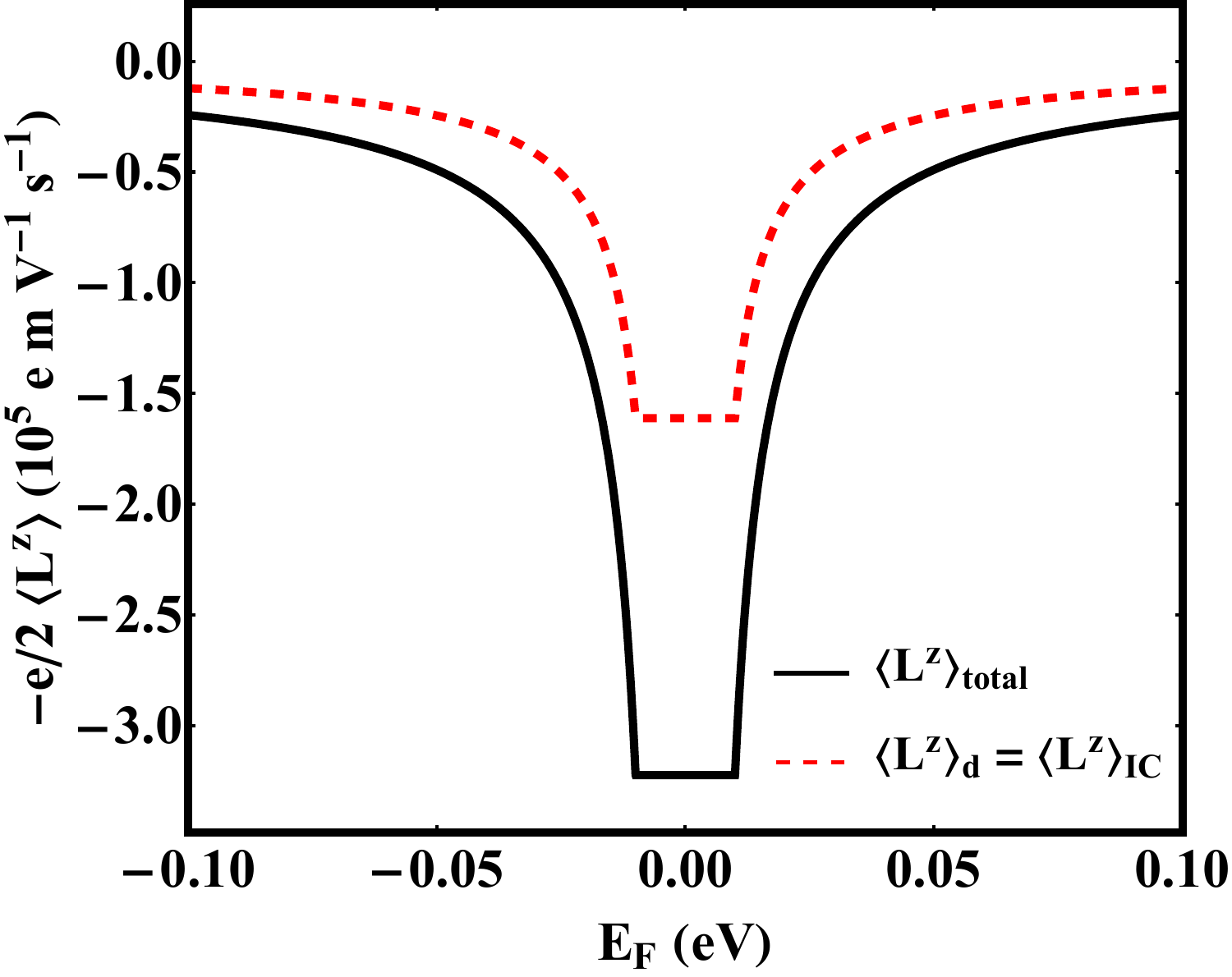}
\caption{\label{Dirac_OAM_plot}
The OMM per unit field and each of its components for our two-band Dirac model. {The dipolar and itinerant curves coincide.} Here we use $\alpha = 4$ eV$\cdot${\AA}, $m = 0.01$ eV, $v_t = 0.2\, \alpha/\hbar$.}
\end{center}
\end{figure}

For the insulating case, $\langle L^z\rangle_{\Sd}$ is calculated to be $\langle L^z\rangle_\Sd =e E^y v_t/12 \pi m$. It is clear that this term is simply $-v_t\langle y \rangle$, which is in agreement with our proposed dipole mechanism for the generation of an OAM. {Since $\langle L^z\rangle_\SIC=\langle L^z\rangle_\Sd$ [Eq.~(\ref{Eq:LIC})], the total OAM in the gap is $eE^yv_t/6\pi m$. For two-band models $\langle L^z\rangle_\SLC$ vanishes identically, so the inter-band OME is exactly $2\langle L^z\rangle_\Sd$. Because $v^{\mu,11}+v^{\mu,22}$ vanishes for a particle-hole-symmetric spectrum, the entire inter-band OME requires particle-hole symmetry breaking, here supplied by the tilt. In the doped regime, $|E_\RF|>m$,
\begin{equation}
\langle L^z\rangle_{\Sd}=\langle L^z\rangle_{\SIC}
={eE^y v_t(m^2+3E_\RF^2)\over48\pi|E_\RF|^3}\quad,\quad
\langle L^z\rangle={eE^y v_t(m^2+3E_\RF^2)\over24\pi|E_\RF|^3}\,.
\label{eq:Dirac_endpoint_result}
\end{equation}}
The nonequilibrium OAM and its components are plotted in Fig.~\ref{Dirac_OAM_plot}.

The massless limit gives the cleanest illustration of the distinction between OEE and inter-band OME. {When $m\rightarrow0$ at fixed $|E_\RF|>0$, the dipolar and itinerant contributions reduce to
\begin{equation}
\langle L^z\rangle_{\Sd}=\langle L^z\rangle_{\SIC}
={eE^y v_t\over16\pi|E_\RF|}\quad,\quad
\langle L^z\rangle={eE^y v_t\over8\pi|E_\RF|}.
\end{equation}}
Moreover the Berry-curvature-like equilibrium OAM of the underlying massless states vanishes. There is therefore no equilibrium OAM texture for the electric field to shift, and consequently no OEE at the Dirac point. The OAM is generated electrically from zero by inter-band, {\it Zitterbewegung}-induced dynamics. In this sense the Dirac point provides a sharp diagnostic example: the OME is not a repopulation of a pre-existing orbital texture, but a genuinely field-induced orbital moment.

This diagnostic simplicity comes with an important symmetry caveat. A single tilted Dirac cone is not normally a complete time-reversal-symmetric system. {In a pair of equally populated time-reversal-related cones the intrinsic inter-band contributions to $L^z$ are valley-odd and cancel.} The response calculated here should therefore be understood as the contribution of one valley, cone, or surface. It could become observable if the two partners are made inequivalent, for example on opposite surfaces of a magnetic topological-insulator heterostructure, in a geometry where one surface dominates the optical or transport response, or in a valley-selective setup in which the two cones are unequally occupied. If the valleys are coupled strongly enough to open a tunnelling gap, however, the massless limit is no longer the appropriate description. Related mesoscopic experiments in graphene quantum dots have also demonstrated the ability to access and manipulate valley-orbital magnetic moments directly, reinforcing the broader experimental relevance of electrically controllable orbital magnetism in low-dimensional systems \cite{PhysRevLett.123.026803-Ensslin&Ihn, Tong_NanoLett_ValleySplitting_2021}.

\begin{figure}[tbp]
\begin{center}
\includegraphics[trim=0cm 0cm 0cm 0cm, clip, width=0.7\columnwidth]{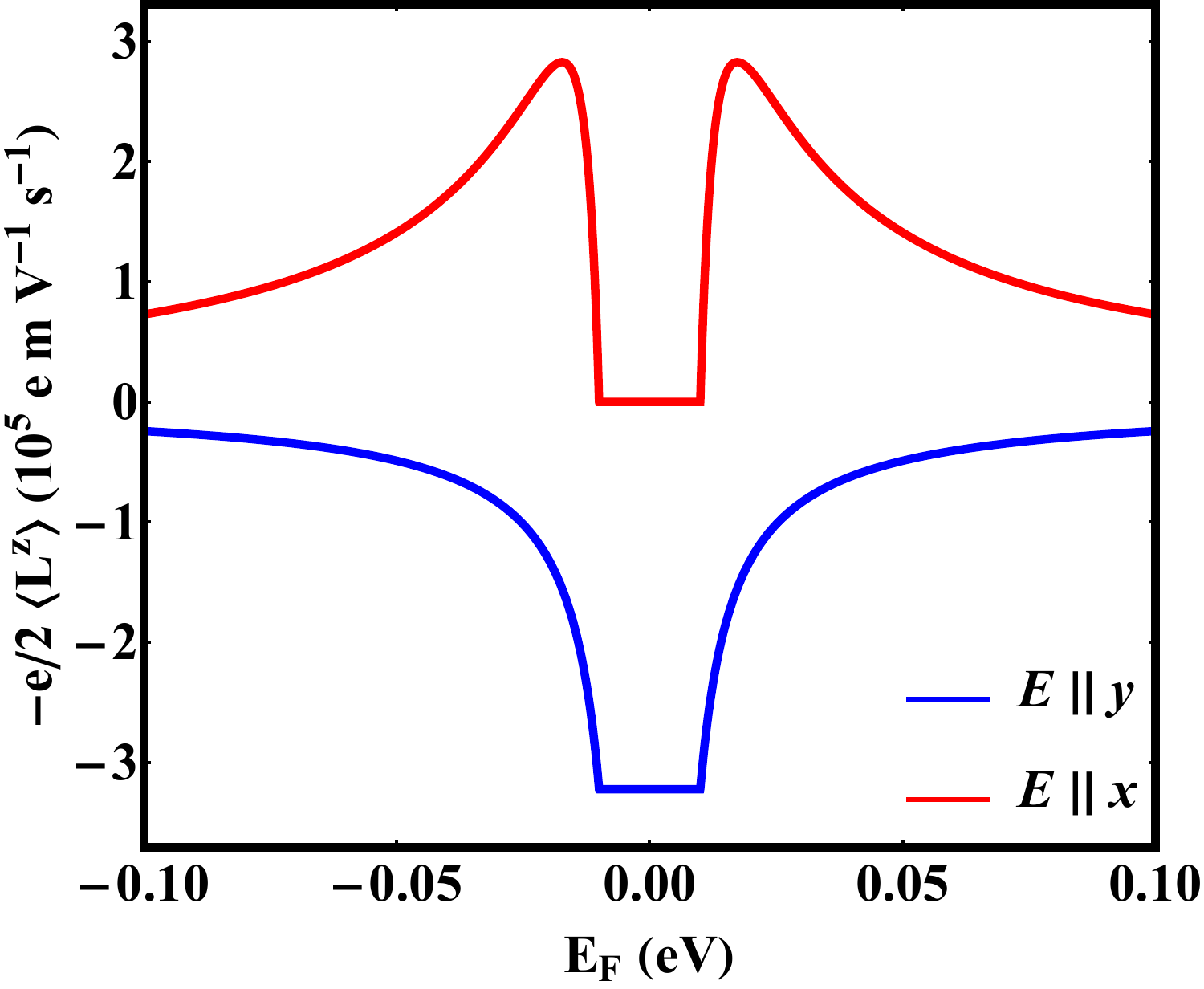}
\caption{\label{OAM_plot_ext}
The {single-cone} inter-band OME ($\vec{E}\parallel\yhat$) and Fermi-surface OEE ($\vec{E}\parallel\xhat$) contributions to the OMM per unit field for our two-band Dirac model. Here we use $\alpha = 4$ eV{\AA}, $m = 0.01$ eV, $v_t = 0.2\, \alpha/\hbar$ and $\tau=0.05$ ps.}
\end{center}
\end{figure}

The OAM density also has a Fermi-surface contribution, that is, the orbital Edelstein effect discussed above. Within the relaxation-time approximation the band-diagonal part of the nonequilibrium density matrix is $\rho^{mm}_\BE = eE^\beta\tau/\hbar\,(\partial f_m/\partial k^{\beta})$ \cite{Interband-Coherence-PRB-2017-Dimi}. {Using Eq.~(\ref{eq:cdod})} gives the Fermi-surface contribution to the effective displacement, $\Xi^{\alpha,mn}=eE^\beta\tau/2\hbar\,\{\CR^\alpha,\partial f/\partial k^\beta\}_{mn}$ with $m\neq n$. Tracing this contribution with the velocity operator yields the OEE plotted in Fig.~\ref{OAM_plot_ext}. For the tilted Dirac cone the OEE is nonzero only for $\BE\parallel\xhat$, whereas the intrinsic inter-band OME is nonzero only for $\BE\parallel\yhat$. This orthogonality provides a direct way to distinguish the two mechanisms. {The valley sum behaves differently in the two channels: under time reversal both the equilibrium OAM texture and the field-induced occupation shift reverse, so their product is valley-even and the OEE contributions add constructively. Thus the OEE survives the valley sum, whereas the intrinsic inter-band contribution cancels for equally populated valleys.} Once the inter-band contribution is isolated, placing the chemical potential in the insulating gap removes the Fermi-surface OEE and makes the dipolar contribution to the OME unambiguous. {The relative magnitudes of the two responses depend on both the chemical potential and the relaxation time; Fig.~\ref{OAM_plot_ext} illustrates the comparison for $\tau=0.05$ ps.} Additional impurity-driven inter-band coherence terms and vertex corrections can modify the Fermi-surface response and are left for future work.

The symmetry of this inter-band OME is therefore different from the conventional Rashba-Edelstein response. In the usual Rashba geometry the electrically generated spin, and in many atom-centred descriptions also the orbital moment, lies in the plane. In the strictly two-dimensional Dirac model considered here the only modern-theory OAM component is out of plane, because in-plane orbital circulation would require motion perpendicular to the plane. The difference is thus not a contradiction but a distinction between response channels: the conventional Edelstein response is a Fermi-surface shift of a pre-existing texture, whereas the OME discussed here is an inter-band coherence response of the modern-theory OAM.

\section{Discussion}
{The expressions above show that the inter-band coherence-induced OME comprises three algebraically distinct contributions. The band-diagonal part of the effective displacement produces a real charge dipole, while the band off-diagonal parts encode local and itinerant circulation; the itinerant term is identical to the dipolar one [Eq.~(\ref{Eq:LIC})], although it arises from the off-diagonal displacement, and the local term requires at least three bands.} In Eq.~(\ref{Eq:Ld}) the dipole is convected by the group velocity and thereby generates OAM. In Eqs.~(\ref{Eq:LLC}) and (\ref{Eq:LIC}) the OAM arises from inter-band matrix elements of the effective displacement and velocity. The three contributions should therefore be understood as different components of the same modern-theory position operator rather than as unrelated mechanisms.

This also clarifies the relation between the OME and OEE terminology. The OEE is often the dominant part of the electrically induced orbital magnetization, but differs from the contribution calculated in Eqs.~(\ref{Eq:Ld})--(\ref{Eq:LIC}). The OEE is the trace of the intra-band correction $n_{\bm E}$ with the equilibrium OAM texture and therefore depends on Fermi-surface occupation and relaxation. The OME, in contrast, is the trace of the inter-band coherence response appearing through $\bm\Xi_{\bm E}$ and contains a Fermi-sea dipolar component even in an insulator. In a generic conductor the measured current-induced OAM density will contain both parts. The primary value of the Dirac model is that symmetry separates them by electric-field orientation, allowing the inter-band part to be discussed without ambiguity.

The connection to quantum geometry is especially transparent for the dipole term. The field-induced displacement in Eq.~(\ref{displacement}) is governed by the normalized quantum metric. This is distinct from the equilibrium OMM, whose familiar expression involves the imaginary part of products of inter-band Berry connections. The OME therefore brings the quantum metric into a linear dc response. Although quantum metric effects are well known in nonlinear optical and superfluid responses \cite{jiang2025revealing,YanaseTAFMphoto,li2024quantum,PhysRevB.108.L201405,doi:10.1126/science.adf1506,wang2023quantum,PhysRevLett.127.277201}, here the metric appears already at first order in a static electric field through the displacement of the Bloch charge distribution. Additionally, the appearance of the quantum metric in the OME suggests that the electrical generation of OAM and the orbital torque could be maximised via quantum-metric engineering.

This displacement also gives a concrete meaning to the {\it Zitterbewegung} picture. The electric field mixes bands virtually, producing an inter-band velocity controlled by the real part of the quantum geometric tensor. The corresponding displacement can be viewed as this quantum-geometric velocity multiplied by an inter-band time scale $\tau_{\textsf Z}\sim\hbar/\Delta\varepsilon$. The response is linear in the dc field, yet bears a formal similarity to rectification because two position operators enter the calculation, one in the perturbation $H_E=e\BE\cdot\Br$ and one in the measured polarization or OAM operator. The density matrix is nevertheless only first order in $H_E$, so the resulting dipole is first order in the field. Expanding the density matrix to second order would give nonlinear responses such as shift and injection currents, while the energy shift $-\Bd_\BE\cdot\BE$ associated with the induced dipole is naturally of second order. Thus the effect is better viewed as the static component of field-induced inter-band coherence, analogous to the inter-band polarization of an insulator but here expressed as an effective displacement within the OAM problem.

The relation between the current theory and the modern theory of polarization is an important insight. The absolute polarization of a periodic crystal depends on a branch choice, while changes in polarization are physical \cite{Vanderbilt_2018,Vanderbilt-polarization-PhysRevB,nunes2001berry,nunes1994real,souza2000polarization,souza2004dynamics}. Our calculation concerns precisely such a change: the electric-field-induced displacement relative to the chosen equilibrium origin. {The origin fixes the homogeneous trace of the equilibrium distributional kernel, whereas the field-induced change is well defined by the endpoint response.} The same structure explains why a finite response can survive in an insulating gap. An insulator has no Fermi-surface shift, but it can have an inter-band polarization.

It is useful to contrast our result with the conventional spin Rashba-Edelstein picture. In a Rashba system the Fermi-surface shift produces an in-plane spin polarization, and in atom-centred orbital descriptions an analogous in-plane orbital polarization may appear. The inter-band OME studied here is qualitatively different. In the two-dimensional Dirac model the modern-theory OAM is out of plane, because the OAM is the cross product of the position and velocity operators, which both lie in the plane. The resulting OME is therefore out of plane and is generated by inter-band coherence rather than by a shifted Fermi surface. In more general three-dimensional systems, or in multi-orbital two-dimensional models where local orbital circulation has additional structure, the symmetry of spin and orbital responses can be more complex. A particularly useful example is a two-dimensional hole gas, where confinement creates quasi-two-dimensional subbands and electric-field-induced transitions between them can generate an in-plane OME even though a strictly two-dimensional Dirac model has only out-of-plane modern-theory OAM \cite{CullenGe2DHoles2026}.

In the Wannier formulation of orbital magnetization and of the OME, itinerant circulation is naturally expressed in terms of boundary currents \cite{thonhauser2011review, malashevich2010theory}. In the Bloch and semiclassical formulations the same contribution is obtained from bulk wave functions without explicitly constructing the boundary. These statements are complementary, not contradictory. The Bloch formalism is formulated for a perfectly periodic infinite crystal, so it cannot display a literal edge current; nevertheless the bulk expression contains the spectral-weight transfer and geometric information that becomes an edge circulation in a finite sample. This is the same logic that underlies the equivalence of the Bloch and Wannier formulations of equilibrium orbital magnetization. The nonzero itinerant contribution in the insulating regime should therefore be viewed as a bulk representation of the same physics that appears as edge transport in a finite Wannier picture.

A finite OME in a gap should likewise not be interpreted as a quantized response. In the tilted Dirac model the insulating response is proportional to $eE^y v_t/(6\pi m)$, and therefore depends on the ratio of the tilt velocity to the gap. More generally, our expressions contain Berry-connection products, velocities, and inter-band energy denominators. Topological structures can enter, but they are not isolated in the form of an integer invariant. Whether a quantized OAM response can be defined under additional symmetry or dimensional constraints remains an open question rather than a consequence of the present formulas.

Previous derivations of the OME provide useful points of comparison for this perspective. Semiclassical expressions for electrically induced OAM were derived in Ref.~\cite{OM-metal-PRB-2021-Xiaocong}, and a fully quantum Bloch-state treatment of the itinerant OME was developed in Ref.~\cite{PhysRevB.107.094106}. Wannier-function expressions were derived in Ref.~\cite{malashevich2010theory}, where the response separates into local and itinerant circulation. Our formulas reproduce the same physical separation while making the density-matrix origin of each term explicit. The remaining differences appear to arise from how the electric-field perturbation and distribution function are treated. In particular, our formulation keeps the diagonal matrix elements of the electric potential and the nonequilibrium density matrix on the same footing, which is essential for isolating the dipole contribution.

\section{Outlook and open questions}

The present work suggests several directions which need to be pursued to achieve a more complete understanding of orbitronic responses. First, the relation between modern-theory OAM and atom-centred OAM out-of-equilibrium remains incomplete. The two approaches agree in some materials and for some equilibrium quantities, but their nonequilibrium relation is not yet understood generally. Second, disorder must be incorporated beyond the relaxation-time estimate used here for the Fermi-surface contribution. The density-matrix framework naturally accommodates anomalous driving terms, side-jump-like terms, and higher-order impurity effects \cite{Interband-Coherence-PRB-2017-Dimi,Hong-OHE-PRL,liu2025quantumOHE}, but their role in the OME remains to be worked out. Third, the connection between bulk orbital currents or itinerant circulation and measurable boundary accumulations is still an open problem, as it is for spin currents. Fourth, orbital-to-spin conversion at realistic interfaces is the key missing ingredient between OAM generation and magnetic switching. Finally, quantum-metric engineering should be explored in realistic materials, including Moir\'e systems, topological antiferromagnets, and heterostructures where symmetry allows the inter-band OME to be separated from the Fermi-surface response.

{A further foundational issue is the real-space origin of nonequilibrium modern-theory OAM. Under a shift of origin by $\bm a$, the raw response changes by $\bm J_{\bm E}\times \bm a$ when the state carries a particle-current density $\bm J_{\bm E}$. Identifying an origin-independent internal OAM without discarding physically relevant itinerant circulation remains an open problem.}

\section{Conclusions}
We have presented a density-matrix perspective on the orbital magneto-electric effect in the modern theory of Bloch-electron OAM. The effective displacement $\bm\Xi$ provides the organizing principle: its trace gives the field-induced dipole, while its contraction with velocity gives the OAM. A static electric field generates a band-diagonal displacement proportional to the normalized quantum metric, giving a linear dc manifestation of inter-band coherence or {\it Zitterbewegung} physics. This displacement contributes directly to the OME and survives in insulating systems. For a tilted massive Dirac cone the inter-band and Fermi-surface orbital responses occur for orthogonal electric-field directions, providing a route to distinguish them experimentally. {The itinerant contribution is identical to the dipolar one, so in a two-band model the dipole accounts for exactly half of the inter-band OME, both in the doped regime and in the insulating gap. In a time-reversal-related pair of equally populated cones the intrinsic inter-band response cancels, whereas the OEE is valley-even and adds constructively.} These results connect the modern theories of orbital magnetization and polarization, and identify the quantum metric as a central design principle for orbitronic materials. {An origin-independent definition of nonequilibrium internal OAM in a current-carrying crystal remains an open problem.}

\section{Acknowledgements} 
This work is supported by the Australian Research Council Discovery Project DP2401062. JHC acknowledges support from an Australian Government Research Training Program (RTP) Scholarship.

\begin{appendix}
\section{The expectation value of dipole operators}
Let $W$ be an operator which commutes with unit cell translations, \ie\ $W$ 
preserves crystal momentum.  Then define the generalized dipole operator $d^\alpha_W\equiv\frac{1}{2}\{r^\alpha,W\}$. The expectation value of this dipole operator is then evaluated as
\begin{align}
\langle d^\alpha_W\rangle&= \half\Tra\big(\rho\,\big\{r^\alpha,W\big\}\big)= 
\half\Tra \big(\big\{r^\alpha,\rho\big\} W\big)\\
&=\Half\sum_{\Bk,\Bq}\sum_{m,n,\ell}\bigg\{\sexpect{m\Bk}{r^\alpha}{n\Bq}\,\sexpect{n\Bq}{\rho}{\ell\Bk}\,\sexpect{\ell\Bk}{W}{m\Bk}\nonumber\\
&\hskip 1.5in+ \sexpect{m\Bk}{\rho}{n\Bq}\,\sexpect{n\Bq}{r^\alpha}{\ell\Bk}\,\sexpect{\ell\Bk}{W}{m\Bk}\bigg\}\ ,\nonumber
\end{align}
where $\rho$ is the single particle density matrix. We then have
\begin{align}
\langle d^\alpha_W\rangle&={1\over 2} \,V\sum_{\ell,m,n}\int\limits_\HOmega\!{d^dk\over 
(2\pi)^d}\int\limits_\HOmega \! d^d q\> \sexpect{n\Bk}{W}{m\Bk}\> \>\Bigg\{
\rho\nd_{\ell\Bq,n\Bk}\>\bigg(\!-i\delta\ns_{m\ell}\,{\pz\over\pz q^\alpha} + \CR^{\alpha,m\ell}_\Bq\bigg)\,\delta(\Bk-\Bq)\nonumber\\
&\hskip 1.8in + \rho\nd_{m\Bk,\ell\Bq}\> \bigg(i\delta\ns_{\ell n}\,
{\pz\over\pz q^\alpha} + \CR^{\alpha,\ell n}_\Bq\bigg)\,\delta(\Bk-\Bq)\Bigg\}\quad,
\end{align}
where $\HOmega$ denotes the first Brillouin zone.  
Integrating by parts, we then have
\begin{equation}
\begin{split}
d^\alpha_W &=\Half\, V\,\sum_{n,m}\int\limits_\HOmega\!{d^dk\over
(2\pi)^d}\>\sexpect{n\Bk}{W}{m\Bk}\>\> \Bigg\{i\lim_{\Bq\to\Bk}{\pz\over
\pz q^\alpha}\big(\rho^{mn}_{\Bq\Bk}-\rho^{mn}_{\Bk\Bq}\big) \\
&\hskip 2.5in +\sum_\ell\Big[\CR^{\alpha,m\ell}_\Bk\,\rho^{\ell n}_\Bk + \rho^{m\ell}_\Bk\,\CR^{\alpha,\ell n}_\Bk\Big]\Bigg\}\quad.
\end{split}
\end{equation}
We may write this as
\begin{equation}
d^\alpha_W=\sum_{m,n}\sum_\Bk\sexpect{n\Bk}{W}{m\Bk}\,
[\CD\rho]^{\alpha,mn}_\Bk\quad ,
\end{equation}
where for any matrix $\Lambda^{mn}_{\Bk\Bk'}=\sexpect{m\Bk}{\Lambda}{n\Bk'}$ 
we define the covariant derivative
\begin{equation}
[\CD\Lambda]^{\alpha,mn}_\Bk=i\lim_{\Bp\to\Bze}\Bigg(
{\pz\Lambda^{mn}_{\Bk_+\hh\Bk_-}\over\pz p^\alpha}
\Bigg)
+\Half\,\big\{\CR^\alpha_\Bk,\Lambda\ns_\Bk\big\}^{mn}\quad,
\label{CovDer}
\end{equation}
with $\Bk_\pm=\Bk\pm\half\Bp$ and $\Lambda^{mn}_\Bk\equiv\lim_{\Bp\to\Bze}
\Lambda^{mn}_{\Bk_+\hh\Bk_-}$.  Note that under a gauge transformation 
$\sket{n\Bk}\to \exp(i\phi\ns_{n\Bk})\,\sket{n\Bk}$ we have
\begin{equation}
[\CD\Lambda]^{\Bk,\alpha}_{m\ell}\to e^{i(\phi\ns_{\ell\Bk}-\phi\ns_{m\Bk})}\,[\CD\Lambda]^{\Bk,\alpha}_{m\ell}\quad ,
\end{equation}
and since $W\ns_{nm}(\Bk) \to e^{i(\phi\ns_{m\Bk}-\phi\ns_{n\Bk})}
W\ns_{nm}(\Bk)$, we have that $d^\alpha_W$ is gauge-invariant.
Under a gauge transformation, the Berry connection transforms as
\begin{equation}
\CR^{\alpha,mn}_\Bk \to \CR^{\alpha,mn}_\Bk - 
{\pz\phi\ns_{m\Bk}\over\pz k^\alpha}\>\delta\ns_{mn}\quad .
\end{equation}
Thus, we have the general result
\begin{equation}\label{Eq:Dipole_Exp}
\langle d^\alpha_A \rangle = \sum_{m,n}\sum_\Bk A^{mn}_\Bk\,\Xi^{\alpha,nm}_\Bk
=\Tra\,(A\ns_\Bk\,\Xi^\alpha_\Bk)\quad ,
\end{equation}
where $\Xi^{\alpha,mn}_\Bk\equiv[\CD\rho]^{\alpha,mn}_{\Bk}$.
Taking $A=1$ yields the expectation value of the position operator
$\langle\Br\rangle = \Tra\,{\BXi}$ in this representation. {For
$\rho_0^{mn}(\Bk,\Bk')=f_m(\Bk)\,\delta_{mn}\,\delta(\Bk-\Bk')$, the complete equilibrium displacement is a two-wave-vector distribution. Its band-diagonal matrix elements are
\begin{align}
\Xi_0^{\alpha,nn}(\Bk,\Bk')={}&{i\over2}
\big[f_n(\Bk)+f_n(\Bk')\big]
{\pz\over\pz k^\alpha}\delta(\Bk-\Bk')\nonumber\\
&+f_n(\Bk)\CR^{\alpha,nn}_\Bk\delta(\Bk-\Bk'),
\label{eq:Xi0d_kernel}
\end{align}
whereas its regular band off-diagonal matrix elements are
\begin{equation}
\Xi_0^{\alpha,mn}(\Bk,\Bk')=\half\CR^{\alpha,mn}_\Bk
\big[f_m(\Bk)+f_n(\Bk)\big]\delta(\Bk-\Bk'),\qquad m\ne n.
\label{eq:Xi0od_kernel}
\end{equation}
The derivative and diagonal-connection pieces in Eq.~(\ref{eq:Xi0d_kernel}) transform together under a Bloch-gauge change. Setting the homogeneous trace $\Tra\,\BXi_0$ to zero fixes the equilibrium centre of mass; it does not imply $\BXi_{0,\Sd}=0$ pointwise. The endpoint construction used below never requires such a pointwise assignment.}

\section{Kinetic equation for $\BXi$}
{
For the response calculation we use the symmetric endpoint continuation of the
homogeneous electric driving term. Let
\begin{equation}
 z=\hbar(\omega+i0^+),\qquad
 \Bk_\pm=\Bk\pm\half\Bp.
\end{equation}
At $\Bp=0$ the usual homogeneous response follows from the linearised kinetic
equation
\[
 {\partial\rho_\BE\over\partial t}+{i\over\hbar}[H_0,\rho_\BE]
 =-{i\over\hbar}[eE^\beta r^\beta,\rho_0],
\]
which, at the retarded frequency $z$, becomes
\[
 [H_0,\rho_\BE(z)]-z\rho_\BE(z)
 =-ieE^\beta(D_\beta\rho_0)_\Bk,
 \qquad D_\beta O\equiv{\partial O\over\partial k^\beta}
 -i[\CR^\beta,O].
\]
The finite-$\Bp$ equation is its symmetric continuation,
\begin{equation}
 H_{0}(\bm k_+)\rho_{\BE,\Bp}(z)-\rho_{\BE,\Bp}(z)H_{0}(\bm k_-)
 -z\rho_{\BE,\Bp}(z)
 =-ieE^\beta(D_\beta\rho_0)_\Bk .
 \label{eq:endpoint_qke_app}
\end{equation}
The matrix on the right-hand side is evaluated in the fixed orbital basis at
the central wave vector and then projected between the Bloch states at
$\Bk_+$ and $\Bk_-$. Equation~(\ref{eq:endpoint_qke_app}) therefore reduces at
$\Bp=0$ to the homogeneous scalar-potential response above.

After parallel transport from both endpoints to $\Bk$, define
\begin{equation}
 \Xi^\alpha_\BE(z)=i\left.
 {\pz\rho_{\BE,\Bp}(z)\over\pz p^\alpha}\right|_{\Bp=0}.
 \label{eq:endpoint_Xi_def}
\end{equation}
In a local band frame this is precisely the covariant first moment in
Eq.~(\ref{CovDer}). Expanding
$H_{0}(\bm k_\pm)=H_0\pm\half p^\alpha\hbar v^\alpha+O(p^2)$ in
Eq.~(\ref{eq:endpoint_qke_app}) gives
\begin{equation}
 [H_0,\Xi^\alpha_\BE(z)]-z\,\Xi^\alpha_\BE(z)
 =-{i\hbar\over2}\{v^\alpha,\rho_\BE(z)\}.
 \label{eq:endpoint_Xi_app}
\end{equation}
There is no first moment of the source because the central matrix
$(D_\beta\rho_0)_\Bk$ is independent of $\Bp$. Thus the distributional
equilibrium kernel in Eq.~(\ref{eq:Xi0d_kernel}) is retained, but no
$D_\beta\Xi_0^\alpha$ term is generated by this endpoint prescription.

The frequency must be kept finite until the first moment has been taken. The
homogeneous response has the Laurent expansion
\begin{equation}
 \rho_\BE(z)={\rho_{-1}\over z}+S_\BE+z\rho_1+O(z^2),
 \label{eq:rho_Laurent}
\end{equation}
where $\rho_{-1}$ is band diagonal, whereas $S_\BE$ and $\rho_1$ are band
off-diagonal to the order required here. Their nonzero elements are
\[
 (\rho_{-1})^{nn}=ieE^\beta\hbar v^{\beta,nn}f'_n,
 \qquad
 (\rho_1)^{mn}=-ieE^\beta\hbar v^{\beta,mn}
 {f_m-f_n\over(\varepsilon_m-\varepsilon_n)^3},\quad m\ne n,
\]
with $S_\BE$ given by Eq.~(\ref{eq:SE_general}). Equivalently, the diagonal
block is $\rho_\BE^{nn}(z)=(\rho_{-1})^{nn}/z$, while for $m\ne n$ the
off-diagonal block is
$\rho_\BE^{mn}(z)=S_\BE^{mn}+z(\rho_1)^{mn}+O(z^2)$.

The two blocks enter Eq.~(\ref{eq:endpoint_Xi_app}) differently. For a
diagonal matrix element its left-hand side is $-z\Xi_\BE^{\alpha,nn}$.
Consequently the regular diagonal displacement is supplied by the
$z\rho_1$ term in Eq.~(\ref{eq:rho_Laurent}), namely
$(\Xi_\BE^{\alpha,nn})_{\rm reg}=(i\hbar/2)
\{v^\alpha,\rho_1\}^{nn}$. This gives
\begin{align}
 (\Xi^{\alpha,nn}_\BE)_{\rm reg}
 ={}&{eE^\beta\hbar^2\over2}\sum_{\ell\ne n}
 \left(v^{\alpha,n\ell}v^{\beta,\ell n}
 +v^{\beta,n\ell}v^{\alpha,\ell n}\right)
 {f_n-f_\ell\over(\varepsilon_n-\varepsilon_\ell)^3}.
 \label{eq:Xi_endpoint_diag}
\end{align}
Its trace is exactly the quantum-metric displacement in
Eq.~(\ref{displacement}); the remaining negative powers of $z$ describe
endpoint accumulation. For $m\ne n$, put
$\Delta_{mn}=\varepsilon_m-\varepsilon_n$ and expand
$(\Delta_{mn}-z)^{-1}=\Delta_{mn}^{-1}+z\Delta_{mn}^{-2}+O(z^2)$.
The regular $S_\BE$ term gives the first line below. The singular diagonal
term $\rho_{-1}/z$ also has a finite residue, generated by the term linear in
$z$ in this denominator expansion, and gives the second line:
\begin{align}
 (\Xi^{\alpha,mn}_\BE)_{\rm reg}
 ={}&-{i\hbar\over2(\varepsilon_m-\varepsilon_n)}
 \{v^\alpha,S_\BE\}^{mn}\nonumber\\
 &\qquad+{eE^\beta\hbar^2\over
 2(\varepsilon_m-\varepsilon_n)^2}v^{\alpha,mn}
 \left(v^{\beta,mm}f'_m+v^{\beta,nn}f'_n\right).
 \label{eq:Xi_endpoint_od}
\end{align}
Thus Eq.~(\ref{eq:Xi_endpoint_diag}) comes from the $z\rho_1$ part of
$\rho_\BE$, while the two lines of Eq.~(\ref{eq:Xi_endpoint_od}) come from
$S_\BE$ and $\rho_{-1}/z$, respectively. The coefficient of the second line
is symmetric under $m\leftrightarrow n$.
Pairing the two ordered-band terms in the OAM trace produces
$2\,\Rep[v^{\nu,nm}v^{\alpha,mn}]$, which is symmetric under
$\alpha\leftrightarrow\nu$ and is therefore annihilated by
$\epsilon_{\lambda\alpha\nu}$ (see Appendix C below). The first line of
Eq.~(\ref{eq:Xi_endpoint_od}) consequently gives Eq.~(\ref{eq:XiE}) and the
complete regular inter-band OME.
}
\section{OAM in an electric field}
The orbital angular momentum can be evaluated as a dipole operator of the form
\begin{equation}
    \left\langle L^\alpha\right\rangle =\eamn \text{Tr}[\frac{1}{2}\{r^\mu,v^
    \nu\}\rho] = \eamn \text{Tr}v^\nu[\frac{1}{2}\{r^\mu,\rho\}]\quad,
\end{equation}
and as such, can be evaluated using Eq.~(\ref{Eq:Dipole_Exp}).
{
Using the regular endpoint coefficients in
Eqs.~(\ref{eq:Xi_endpoint_diag}) and (\ref{eq:Xi_endpoint_od}), the
electric-field-induced OAM is
\begin{align}
 \langle L^\alpha\rangle_\BE
 ={}&\eps_{\alpha\mu\nu}\sum_{\Bk,n}
 v^{\nu,nn}_\Bk(\Xi^{\mu,nn}_\BE)_{\rm reg}\nonumber\\
 &\qquad-{i\hbar\over2}\eps_{\alpha\mu\nu}\sum_\Bk\sum_{m\ne n}
 {v^{\nu,nm}_\Bk\{v^\mu,S_\BE\}^{mn}
 \over\varepsilon_m-\varepsilon_n}\nonumber\\
 \equiv{}&\LASd+\LAIC+\LALC .
 \label{OAME}
\end{align}
The first line gives Eq.~(\ref{Eq:Ld}). To separate the second line, write
\begin{align}
 \{v^\mu,S_\BE\}^{mn}
 ={}&(v^{\mu,mm}+v^{\mu,nn})S_\BE^{mn}\nonumber\\
 &\qquad+\sum_{\ell\notin\{m,n\}}
 \left(v^{\mu,m\ell}S_\BE^{\ell n}
 +S_\BE^{m\ell}v^{\mu,\ell n}\right).
 \label{eq:anticommutator_split}
\end{align}
The repeated-band term in the first line of
Eq.~(\ref{eq:anticommutator_split}) gives the itinerant contribution
$\LAIC$. Substituting Eq.~(\ref{eq:SE_general}) gives Eq.~(\ref{Eq:LIC});
interchanging the dummy indices and the two antisymmetric OAM indices shows
directly that $\LAIC=\LASd$. The all-distinct term in the second line gives
the genuine three-band local-circulation expression in Eq.~(\ref{Eq:LLC}).
It vanishes identically when only two bands are present.
}
\section{Separation into itinerant and local circulation}
{The itinerant/local terminology is motivated by the decomposition of the position operator below, although, as discussed at the end of this appendix, the separation used in the main text is algebraic.} Whereas local circulation refers to orbital motion within the unit cell, itinerant circulation refers to orbital motion of the electron through the lattice. Now, starting with the matrix element of the position operator between two Bloch states $\sexpect{n\Bk}{r^\alpha}{l\Bq}$, this can then be expressed as
\begin{equation}
\begin{aligned}
\sexpect{n\Bk}{r^\alpha}{l\Bq} &= \int\! dV\, e^{i(\Bq-\Bk)\cdot\boldsymbol{r}} u^*_{n \Bk}\, r^\alpha\, u\ns_{l\Bq}\\
&= \sum_{\BR}\int\limits_\Omega\! dv\, e^{i(\Bq-\Bk)\cdot(\BR+\Bx)}\, u^*_{n \Bk} (R^\alpha + x^\alpha)\, u\ns_{l\Bq}\quad.
\end{aligned}
\end{equation}
Here the volume integral has been separated into a sum of integrals over each unit cell, where $\boldsymbol{r}=\boldsymbol{R}+\boldsymbol{x}$, $\boldsymbol{R}$ is the position of the cell center and $\boldsymbol{x}$ is the position within the cell. Now
\begin{equation}
\begin{split}
    \sexpect{n\Bk}{r^\alpha}{l\Bq} &= \sum_\BR e^{i(\Bq-\Bk)\cdot\BR}  R^\alpha \int\limits_\Omega\! dv\, e^{i(\Bq-\Bk)\cdot\Bx} u^*_{n \Bk} u\ns_{l\Bq} \\
&\hskip1.2in + \sum_\BR e^{i(\Bq-\Bk)\cdot\BR} \int\limits_\Omega\! dv\, e^{i(\Bq-\Bk)\cdot\Bx} u^*_{n \Bk}\, x^\alpha\, u\ns_{l\Bq} \quad,
\end{split}
\end{equation}
here, we can note that the first term represents the position within the lattice and the second term represents the position within the unit cell. Now, we can finish evaluating these terms, the first term becomes
\begin{align}
\sum_\BR e^{i(\Bq-\Bk)\cdot\boldsymbol{R}}  R^\alpha \int_{\Omega} dv\, e^{i(\Bq-\Bk)\cdot\Bx} u^*_{n \Bk} u_{l\Bq}
&=i \sum_\BR \pd{}{k^\alpha}  e^{i(\Bq-\Bk)\cdot\BR} \int_{\Omega} dv\, e^{i(\Bq-\Bk)\cdot\Bx} u^*_{n \Bk} u\ns_{l\Bq}\nonumber\\
&=i\,\pd{}{k^\alpha}\, \delta_{nl}\,\delta_{\Bk\Bq}\quad,
\end{align}
and the second term
\begin{align}
\sum_\BR e^{i(\Bq-\Bk)\cdot\BR} \int\limits_\Omega\! dv\, e^{i(\Bq-\Bk)\cdot\Bx} u^*_{n \Bk}\, x^\alpha\, u\ns_{l\Bq}
&=- i \sum_\BR e^{i(\Bq-\Bk)\cdot\BR} \int\limits_\Omega\! dv\, 
e^{i(\Bq-\Bk)\cdot\Bx}\, \pd{u^*_{n \Bk}}{k^\alpha}\, u\ns_{l\Bq}\nonumber\\
&=\CR^{\alpha,nl}_{\Bk}\, \delta_{\Bk\Bq}\quad.
\end{align}
{
The differential and Berry-connection pieces of Eq.~(\ref{CovDer}) therefore
encode, respectively, motion through the lattice and position within the unit
cell. In the endpoint response, however, this separation must be made only
after the complete finite-$\Bp$ product has been formed. For $m\ne n$, the
regular OAM-producing displacement is proportional to
\begin{align}
 \{v^\mu,S_\BE\}^{mn}
 ={}&(v^{\mu,mm}+v^{\mu,nn})S_\BE^{mn}\nonumber\\
 &\qquad+\sum_{\ell\notin\{m,n\}}
 \left(v^{\mu,m\ell}S_\BE^{\ell n}
 +S_\BE^{m\ell}v^{\mu,\ell n}\right).
\end{align}
The repeated-band first line is the term conventionally grouped as itinerant
circulation, whereas the all-distinct second line is the local-circulation
term. The band-diagonal endpoint coefficient gives the convected dipole.
These are useful algebraic groupings and retain the notation of
Table~\ref{tab:OAMcomp}; the complete OAM is independent of how the
off-diagonal sector is partitioned. More generally, in a current-carrying
state the raw OAM depends on the real-space origin, as noted in the Outlook.
}
\end{appendix}

\bibliographystyle{tfq}
\bibliography{OAM}
\end{document}